\documentclass[twoside]{article}
\usepackage{qic}

\usepackage{amsmath}
\usepackage{amssymb}
\usepackage{braket}
\usepackage{color}

\newtheorem{thm}{Theorem}

\begin{document}
\setlength{\textheight}{8.0truein}    

\runninghead{A localized quantum walk with a gap in distribution}
            {T. Machida}

\normalsize\textlineskip
\thispagestyle{empty}
\setcounter{page}{1}

\vspace*{0.88truein}

\alphfootnote

\fpage{1}

\centerline{\bf
A localized quantum walk with a gap in distribution}
\vspace*{0.37truein}
\centerline{\footnotesize
Takuya Machida}
\vspace*{0.015truein}
\centerline{\footnotesize\it College of Industrial Technology, Nihon University,}
\baselineskip=10pt
\centerline{\footnotesize\it Narashino, Chiba 275-8576, Japan}
\vspace*{0.225truein}

\vspace*{0.21truein}

\abstracts{
Quantum walks behave differently from what we expect and their probability distributions have unique structures.
They have localization, singularities, a gap, and so on.
Those features have been discovered from the view point of mathematics and reported as limit theorems.
In this paper we focus on a time-dependent three-state quantum walk on the line and demonstrate a limit distribution.
Three coin states at each position are iteratively updated by a coin-flip operator and a position-shift operator. 
As the result of the evolution, we end up to observe both localization and a gap in the limit distribution.    
}{}{}

\vspace*{10pt}

\vspace*{3pt}

\vspace*{1pt}\textlineskip    

\bibliographystyle{qic}

\section{Introduction}
\label{sec:introduction}
Quantum walks are quantum counterparts of random walks from the view point of mathematics.
Since they are also considered to be quantum algorithms in themselves, studies of quantum walks have been associated to the field of quantum information.
The behavior of quantum walks is very different from that of random walks in probability distribution and the differences are useful to build some quantum search algorithms which are quadratically faster than classical ones~\cite{Venegas-Andraca2008,Venegas-Andraca2012}.
Apart from quantum algorithms, we take care of a quantum walk as a mathematical model in this paper.
The study of quantum walks started to get attention from mathematics around 2000 and the interesting behavior of them has been reported.  
One would say that long-time limit theorems are representative results for quantum walks.
As of now, it is hard to know the behavior of quantum walks statistically after the walkers have repeated their evolutions a lot of times, because there are difficulties in experiments to perform the evolutions in physical systems maintaining their quantum states.
Long-time limit theorems tell us how the walkers approximately act, and they are helpful to understand quantum walks after the walkers have got updated many times.
The first limit theorem was reported for a probability distribution in 2002 and it described a convergence in distribution in a rescaled space by time~\cite{Konno2002a}.
A density function obtained in the theorem completely captured some features of the probability distribution at finite long-time.
One of the remarkable features was that the probability distribution was spreading in proportion to time as the walker was iterating its time evolution.
Seeing the limit theorem, we also make it clear that the probability distribution steeply goes up at two points.
That fact was pictured as two singularities in the limit density function.
Each point which gave a singularity to the limit density function was determined by a time evolution operator known as coin-flip operator.
The exact locations of the singular points were hard to find from numerical experiments and those points were figured out for the first time after the limit theorem was derived.
Such a mathematical analysis helps us understand quantum walks.
Up until now since the first limit theorem was born, a lot of limit theorems have been proved~\cite{Venegas-Andraca2012}.

In this paper we analyze a time-dependent three-state quantum walk which gets updated according to a 3-periodic evolution.
A motivation for the walk comes from a past study about a 3-period time-dependent two-state quantum walk~\cite{GrunbaumMachida2015}.
The two-state walk followed a 3-periodic evolution and resulted in a probability distribution with a gap.
Distributions with a gap had never been discovered in the field of quantum walks before the paper was published.
That discovery naturally shifts our focuses to three-state walks because it was reported that a three-state walk localized unlike the two-state walk~\cite{InuiKonnoSegawa2005,vStefavnakBezdvekovaJex2014,Machida2015}.
We try to find a three-state walk whose distribution features both localization and a gap, and verify long-time behavior of the walker by deriving a limit theorem.

We start off with the definition of a time-dependent three-state quantum walk in the following section.
The walker is updating itself according to a 3-periodic rule.
In Sec.~\ref{sec:limit_th}, a limit theorem is demonstrated with its proof.
A convergence in distribution in a rescaled space is shown in the limit theorem and it informs about both localization and ballistic spreading of the walker.
Particularly, the ballistic spreading part of the limit density function is exactly computed.
The summary is given in the final section where we will mention what the limit theorem tells us.

\section{Definition of a 3-period time-dependent three-state quantum walk on the line}
\label{sec:definition}
A time-dependent three-state quantum walk presented in this paper is defined in a Hilbert space which is built from a position Hilbert space and a coin Hilbert space.
The location of the walker is determined on the line $\mathbb{Z}=\left\{0,\pm 1,\pm 2,\ldots\right\}$ and its position $x\in\mathbb{Z}$ is mapped to a vector $\ket{x}$ in a Hilbert space $\mathcal{H}_p$ whose orthogonal normalized basis $\left\{\ket{x}:\,x\in\mathbb{Z}\right\}$.
The walker at each position is assumed to be in a superposition state of three states to which the numbers $-1,0$, and 1 are assigned, respectively.
The three states are also called coin states.
The superposition state is expressed in a Hilbert space $\mathcal{H}_c$ whose orthogonal normalized basis $\left\{\ket{-1}, \ket{0}, \ket{1}\right\}$.
The vector $\ket{j}\,(j\in\left\{-1,0,1\right\})$ in the Hilbert space $\mathcal{H}_c$ is considered to be a counterpart to the $j$-th coin state of the walker.
Totally, the superposition state of the coin states all over the line turns out to be descried in the tensor Hilbert space $\mathcal{H}_p\otimes\mathcal{H}_c$, such as $\sum_{x\in\mathbb{Z}}\ket{x}\otimes\left(\rho_{-1}(x)\ket{-1}+\rho_0(x)\ket{0}+\rho_1(x)\ket{1}\right)$ with complex numbers $\rho_{-1}(x),\rho_0(x)$, and $\rho_1(x)$, which means that the superposition state of the coin states at position $x$ is of the form $\rho_{-1}(x)\ket{-1}+\rho_0(x)\ket{0}+\rho_1(x)\ket{1}$.

Let us represent the superposition state all over the line at time $t\,(\in\left\{0,1,2,\ldots\right\})$ as the form $\ket{\Psi_t}=\sum_{x\in\mathbb{Z}}\ket{x}\otimes\ket{\psi_{t}(x)}\in\mathcal{H}_p\otimes\mathcal{H}_c$.
The time evolution takes place with a coin-flip operator and a position-shift operator.
In this study we use a parameterized unitary operator
\begin{align}
 C=&-\frac{1+c}{2}\ket{-1}\bra{-1}+\frac{s}{\sqrt{2}}\ket{-1}\bra{0}+\frac{1-c}{2}\ket{-1}\bra{1}+\frac{s}{\sqrt{2}}\ket{0}\bra{-1}+c\ket{0}\bra{0}+\frac{s}{\sqrt{2}}\ket{0}\bra{1}\nonumber\\
 &+\frac{1-c}{2}\ket{1}\bra{-1}+\frac{s}{\sqrt{2}}\ket{1}\bra{0}-\frac{1+c}{2}\ket{1}\bra{1},\label{eq:coin-flip_operator}
\end{align}
to transit the superposition state of the coin states $\ket{-1}, \ket{0}$, and $\ket{1}$ at each position, where the letters $c$ and $s$ are short for $\cos\theta$ and $\sin\theta\,(\theta\in [0,2\pi))$, respectively.
The variable $\theta$ denotes the parameter of the operator $C$ through this paper.
Particularly, if we let $\theta=\arccos(-1/3)$, then a Grover operator
\begin{align}
 C=&-\frac{1}{3}\ket{-1}\bra{-1}+\frac{2}{3}\ket{-1}\bra{0}+\frac{2}{3}\ket{-1}\bra{1}+\frac{2}{3}\ket{0}\bra{-1}-\frac{1}{3}\ket{0}\bra{0}+\frac{2}{3}\ket{0}\bra{1}\nonumber\\
 &+\frac{2}{3}\ket{1}\bra{-1}+\frac{2}{3}\ket{1}\bra{0}-\frac{1}{3}\ket{1}\bra{1},\label{eq:grover_coin}
\end{align}
shows up.
The Grover operator was also picked in Inui et al.~\cite{InuiKonnoSegawa2005} and its extension formed in Eq.~\eqref{eq:coin-flip_operator} was employed in Machida~\cite{Machida2015}.
Changing the parameter appropriately in Eq.~\eqref{eq:coin-flip_operator}, we obtain an operator used in {\v{S}}tefa{\v{n}}{\'a}k et al.~\cite{vStefavnakBezdvekovaJex2014}.
We can, therefore, recognize the difference between the limit theorem presented in this paper and those shown already in the past studies, and that is a reason that the form in Eq.~\eqref{eq:coin-flip_operator} has been assigned to the operator $C$.
Using the parameterized operator in Eq.~\eqref{eq:coin-flip_operator}, we opt to generate the superposition state at time $t+1$ from that at time $t$,
\begin{equation}
 \ket{\Psi_{t+1}}=\left\{\begin{array}{ll}
		   \tilde{S}\tilde{C}\ket{\Psi_t}& (t=0,1 \mod 3)\\[1mm]
			  \tilde{S}\ket{\Psi_t}& (t=2 \mod 3)
			 \end{array}\right.,
\label{eq:time-evolution}
\end{equation}
where
\begin{align}
 \tilde{C}=&\sum_{x\in\mathbb{Z}}\ket{x}\bra{x}\otimes C,\\
 \tilde{S}=&\sum_{x\in\mathbb{Z}}\ket{x-1}\bra{x}\otimes\ket{-1}\bra{-1}+\ket{x}\bra{x}\otimes\ket{0}\bra{0}+\ket{x+1}\bra{x}\otimes\ket{1}\bra{1}.
\end{align}
The superposition state of the coin states at each position gets updated by a coin-flip operator $\tilde{C}$.
On the other hand, the operator $\tilde{S}$ plays a role to shift the walker with the coin state $\ket{-1}$ (resp. $\ket{0}, \ket{1}$) from position $x$ to $x-1$ (resp. $x, x+1$) without changing the coin state.
It is, hence, called position-shift operator.
As shown in Eq.~\eqref{eq:time-evolution}, the superposition state does not get changed at time $t\in\left\{2,5,8,\ldots\right\}$ (i.e. $t=2 \mod 3$) before the position states get changed at the time.
If we substitute either $0$ or $\pi$ to the variable $\theta$, then the coin-flip operator takes the form
\begin{equation}
 C=-\ket{-1}\bra{-1}+\ket{0}\bra{0}-\ket{1}\bra{1},\quad \ket{-1}\bra{1}-\ket{0}\bra{0}+\ket{1}\bra{-1},
\end{equation}
respectively.
These special operators perform trivial behavior of the walker.
For this reason, the delicate analysis of the quantum walk operated by the coin-flip operator with such values is not demanded and we assume $\theta\neq 0,\pi$ after this point.

\begin{figure}[h]
\begin{center}
 \begin{minipage}{70mm}
  \begin{center}
   \includegraphics[scale=0.3]{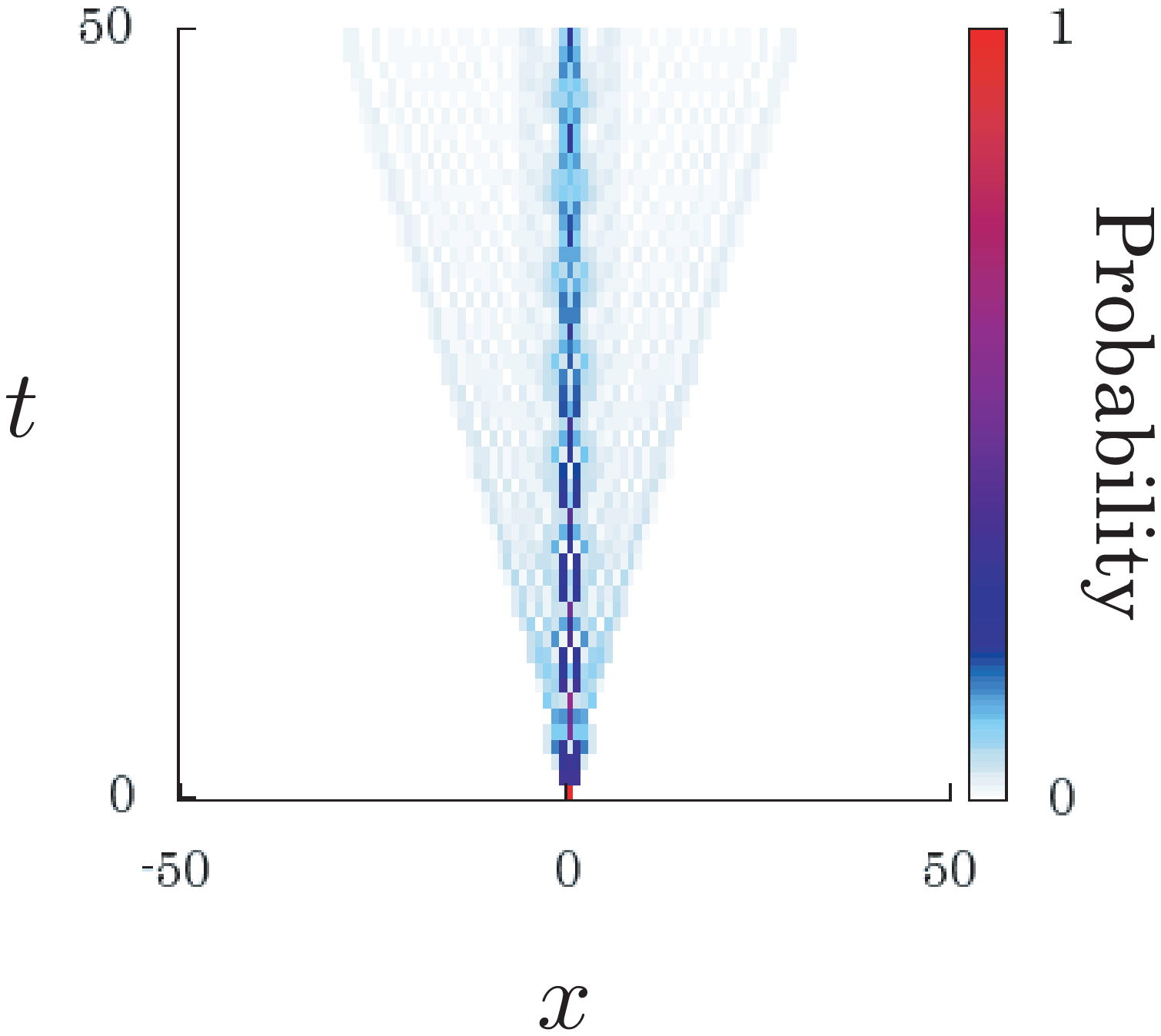}\\[2mm]
  (a) $\theta=\arccos(-1/3)$
  \end{center}
 \end{minipage}
 \begin{minipage}{70mm}
  \begin{center}
   \includegraphics[scale=0.3]{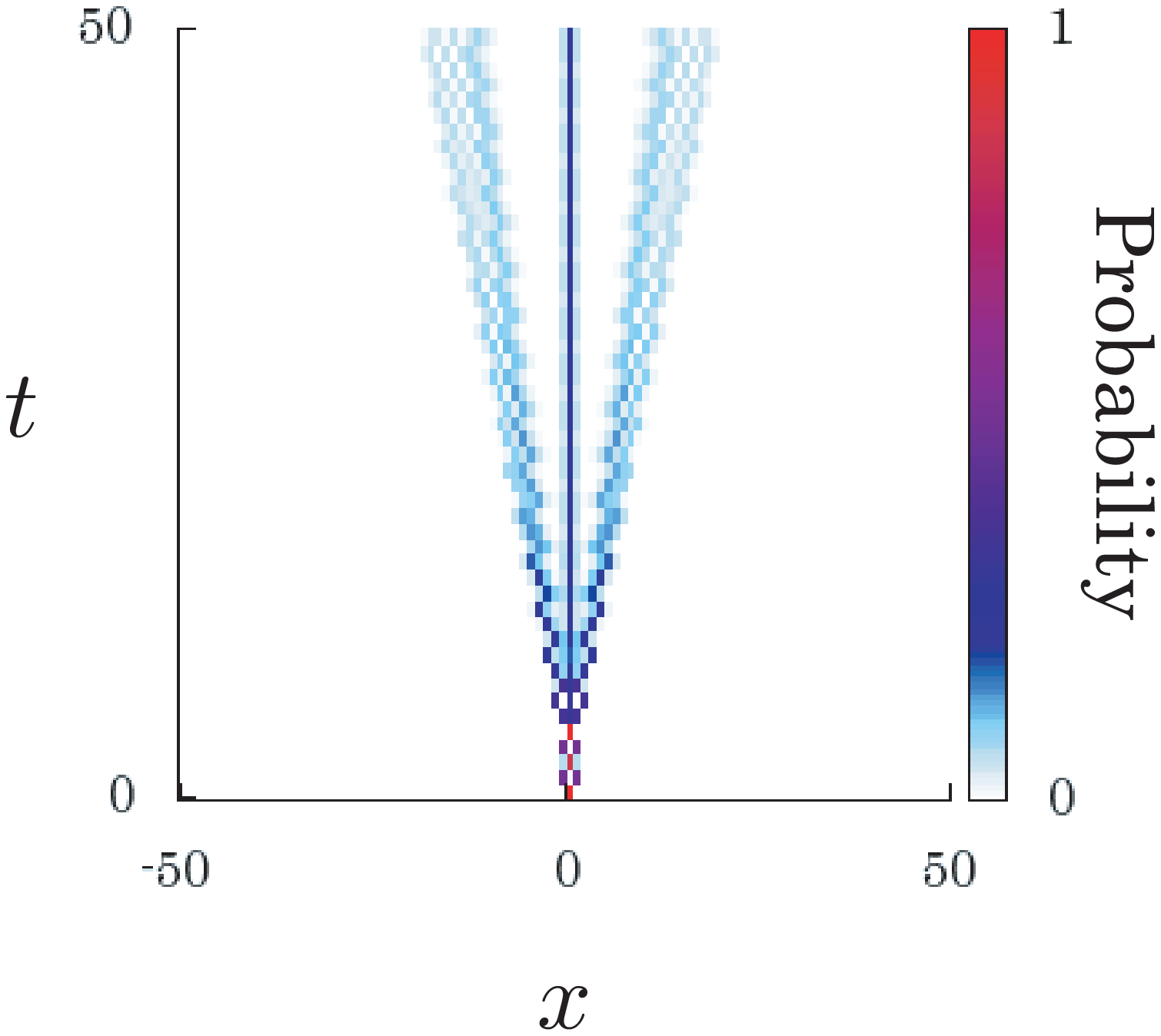}\\[2mm]
  (b) $\theta=5\pi/6$
  \end{center}
 \end{minipage}
\vspace{5mm}
\fcaption{Time evolution of the probability distribution $\mathbb{P}(X_t=x)$ in the case of $\alpha=\beta=\gamma=1/\sqrt{3}$ : The walker starts with the initial state given in Eq.~\eqref{eq:initial_state}. We see localization around the origin and ballistic spreading in proportion to time $t$ in these pictures.}
\label{fig:time-probability}
\end{center}
\end{figure}

\begin{figure}[h]
\begin{center}
 \begin{minipage}{70mm}
  \begin{center}
  \includegraphics[scale=0.3]{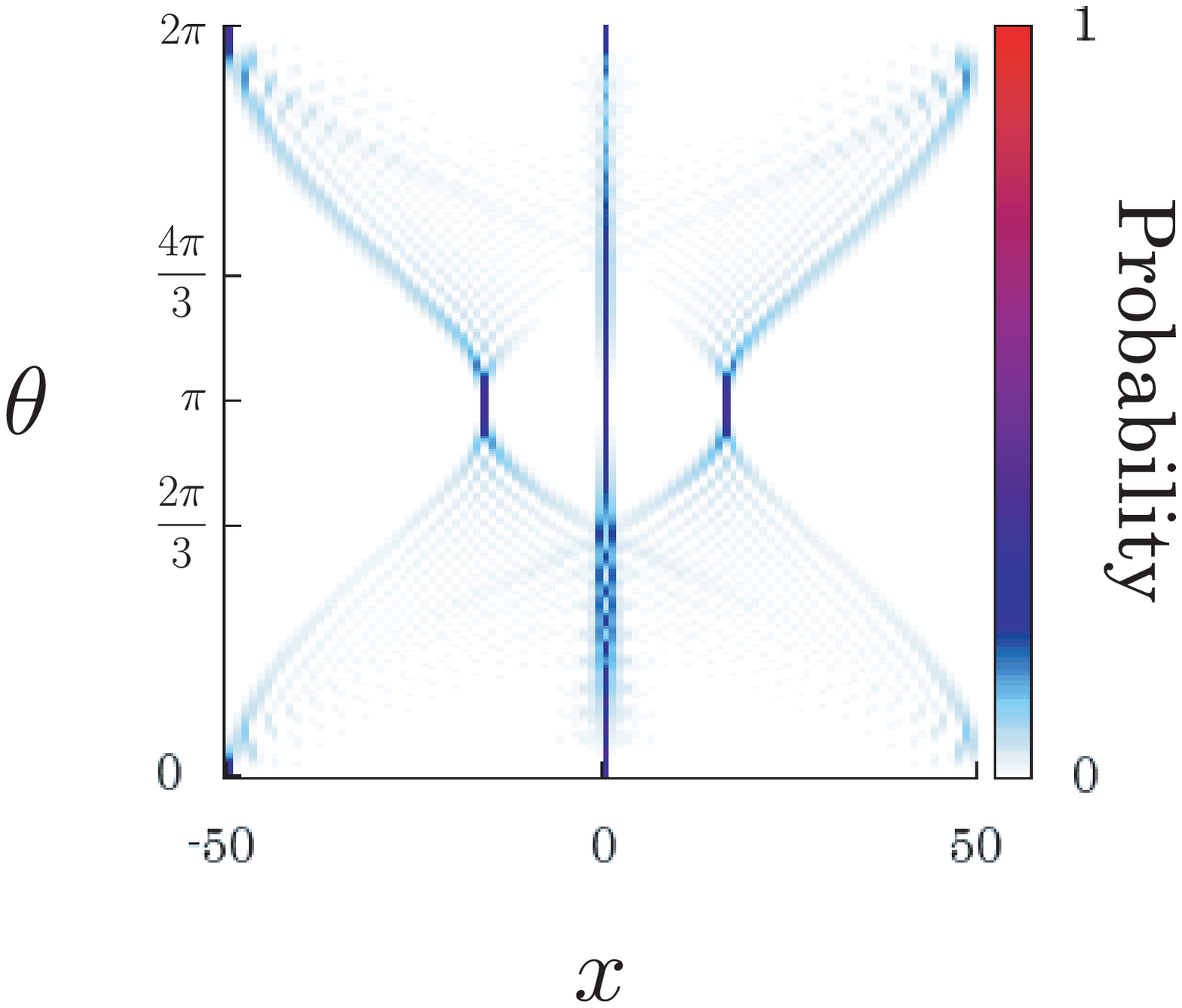}\\[2mm]
  (a) $\alpha=1/\sqrt{3},\,\beta=1/\sqrt{3},\,\gamma=1/\sqrt{3}$
  \end{center}
 \end{minipage}
 \begin{minipage}{70mm}
  \begin{center}
   \includegraphics[scale=0.3]{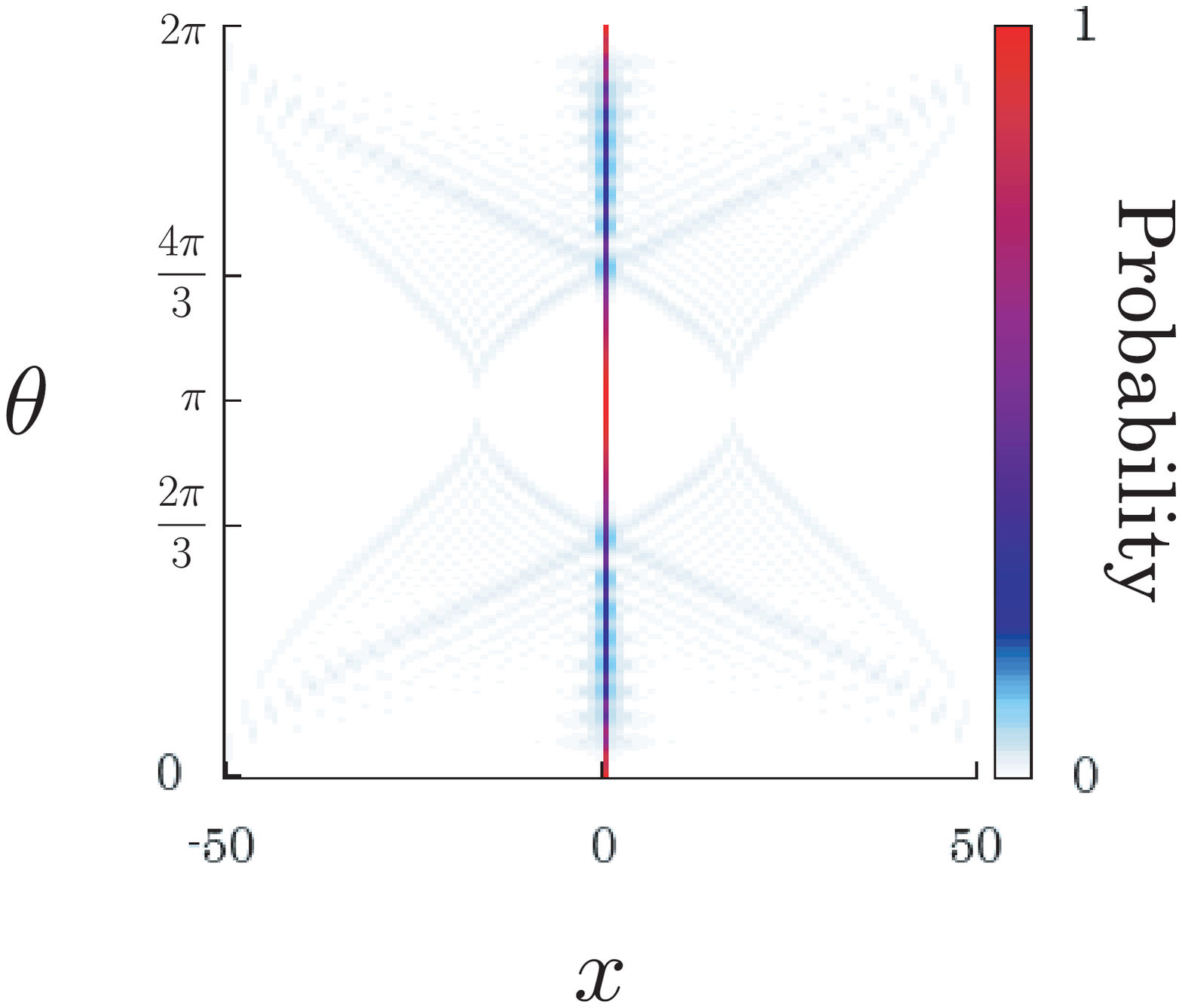}\\[2mm]
  (b) $\alpha=0,\,\beta=1,\,\gamma=0$
  \end{center}
 \end{minipage}
\vspace{5mm}
\fcaption{Relation between the probability distribution at time $50$ and the parameter $\theta$ which determines the coin-flip operator $C$ : We observe a gap in the probability distribution for $\theta\in (2\pi/3,4\pi/3)$ in both pictures.}
\label{fig:theta-probability}
\end{center}
\end{figure}

Under the configuration written above, our focuses are put on a likelihood with regard to the position of the walker.
The walker is observed at a location on the line and its position is determined at random according to a probability law.
The law is based on a notion of quantum physics and the probability that the walker is observed at position $x$ at time $t$ is given by
\begin{equation}
 \mathbb{P}(X_t=x)=\bra{\Psi_t}\biggl\{\ket{x}\bra{x}\otimes \Bigl(\ket{-1}\bra{-1}+\ket{0}\bra{0}+\ket{1}\bra{1}\Bigr)\biggr\}\ket{\Psi_t},\label{eq:prob_dist}
\end{equation}
assuming $\braket{\Psi_0|\Psi_0}=1$.
The letter $X_t$ means the position of the walker at time $t$.
In this study we suppose that the initial state takes the following representation.
Given complex numbers $\alpha, \beta$, and $\gamma\in\mathbb{C}$ subject to the constraint $|\alpha|^2+|\beta|^2+|\gamma|^2=1$, the walker holds the superposition state $\ket{0}\otimes\bigl(\alpha\ket{-1}+\beta\ket{0}+\gamma\ket{1}\bigr)$ at time $0$, that is,
\begin{equation}
 \ket{\Psi_0}=\ket{0}\otimes\Bigl(\alpha\ket{-1}+\beta\ket{0}+\gamma\ket{1}\Bigr),\label{eq:initial_state}
\end{equation}
where $\mathbb{C}$ denotes the set of complex numbers.
Initial states like Eq.~\eqref{eq:initial_state} are normally configured for quantum walks to compare with random walks.
The superposition state defined in Eq.~\eqref{eq:initial_state} actually gives the initial probability distribution $\mathbb{P}(X_0=0)=1$ and $\mathbb{P}(X_0=x)=0\,(x\neq 0)$ which is also commonly used for the study of random walks starting off at the origin at time $0$. 
Interested in the probability distribution given in Eq.~\eqref{eq:prob_dist}, we will show a limit theorem for it in the next section.
The theorem asserts that the walker has both localization and ballistic behavior.
A glance at Fig.~\ref{fig:time-probability} also illustrates localization and ballistic spreading of the walker as time $t$ is going up. 
It seems that the probability distribution localizes around the origin and diffuses in proportion to time $t$.
In particular, we see a gap arising in Fig.~\ref{fig:time-probability}--(b).
Figure~\ref{fig:theta-probability} shows how the probability distribution at time $50$ depends on the parameter $\theta$.
One can guess that a gap occurs in the probability distribution for $\theta\in (2\pi/3,4\pi/3)$.
We should note that the value $5\pi/6$ which has been used for the parameter $\theta$ in Fig.~\ref{fig:time-probability}--(b) is in the interval.
In Sec.~\ref{sec:limit_th}, the limit theorem will confirm these expected phenomena.

\section{Long-time behavior described by a limit theorem}
\label{sec:limit_th}
\begin{figure}[h]
\begin{center}
 \begin{minipage}{70mm}
  \begin{center}
   \includegraphics[scale=0.5]{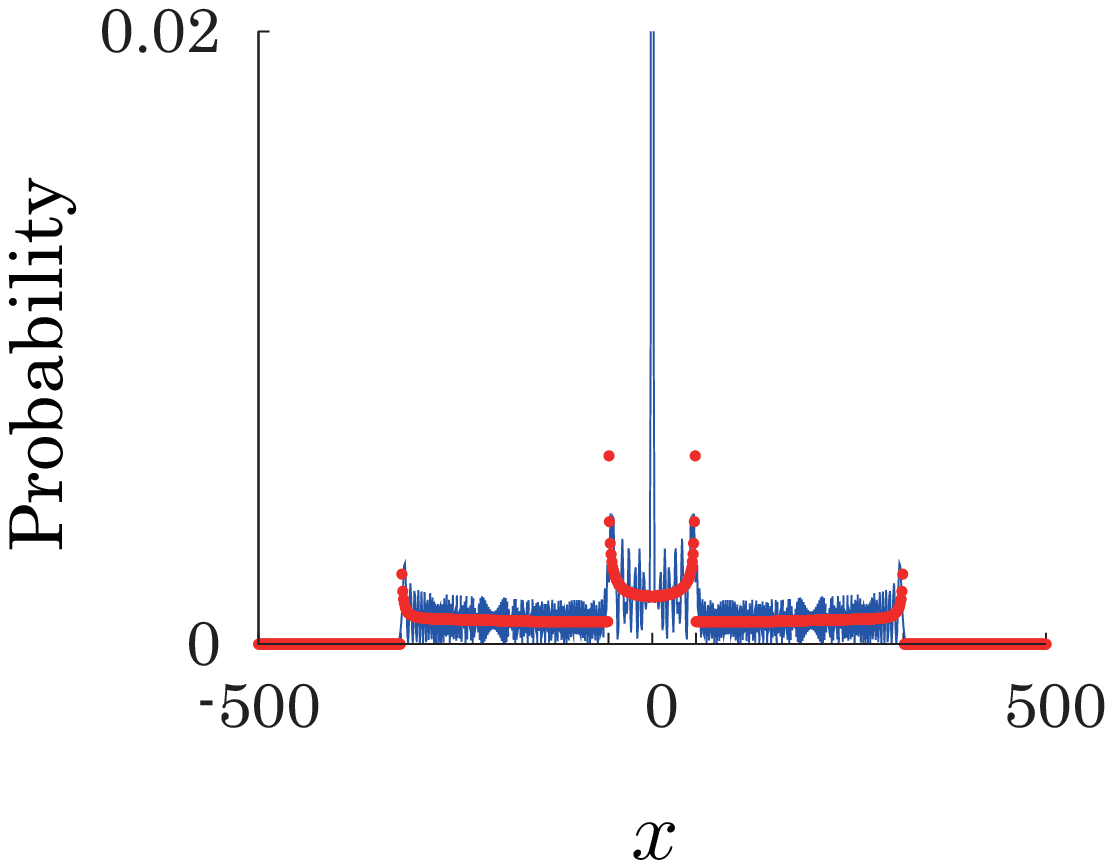}\\[2mm]
  (a) $\theta=\arccos(-1/3)$
  \end{center}
 \end{minipage}
 \begin{minipage}{70mm}
  \begin{center}
   \includegraphics[scale=0.5]{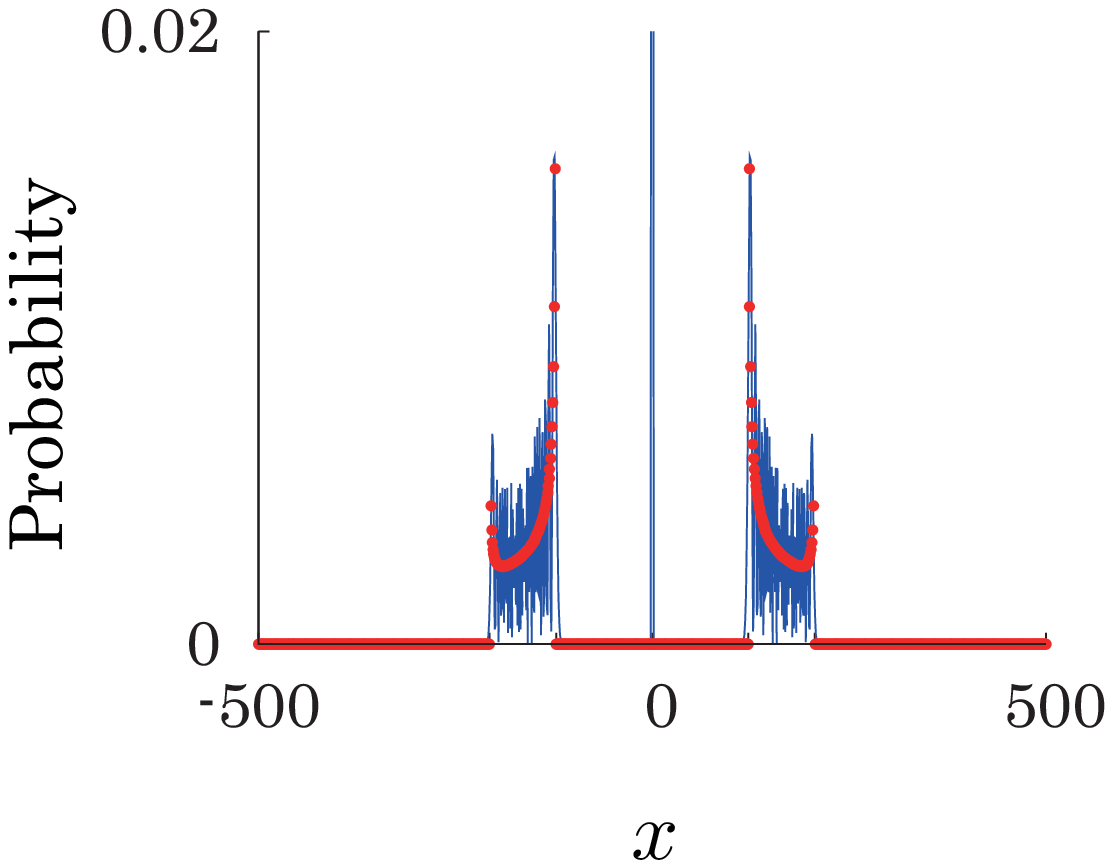}\\[2mm]
  (b) $\theta=5\pi/6$
  \end{center}
 \end{minipage}
\vspace{5mm}
\fcaption{The probability distribution at time $500$ (blue line) and the function on the right side of Eq.~\eqref{eq:approximation} (red point), in the case of $\alpha=\beta=\gamma=1/\sqrt{3}$ : The approximate function catches the ballistic spreading part of the probability distribution. We also observe a gap in Fig.~(b) when the value of the parameter $\theta$ is equal to $5\pi/6$.}
\label{fig:limit}
\end{center}
\end{figure}

Long-time limit theorems are very useful to understand the behavior of quantum walks after their evolutions have been repeated a lot of times.
We see a limit theorem for the 3-period time-dependent three-state quantum walk in this section.
The limit theorem tells us how and where the walker asymptotically distributes after time $t$ goes enough up.
As the result, we can also exactly estimate how the distribution depends on the initial parameters $\alpha, \beta,\gamma$, and the coin parameter $\theta$.  
The asymptotic behavior will be actually described in Eq.~\eqref{eq:approximation} and visualized in Fig.~\ref{fig:limit} later, which convinces us that the approximation matches probability distribution of the quantum walk very well.  
Since there are difficulties to implement quantum walks on physical systems, it is hard to know the behavior of quantum walks in statistics now.
Long-time limit theorems give an important contribution to quantum walks.

Limit theorems for a three-state quantum walk on the line were reported for the first time in 2005 by Inui et al.~\cite{InuiKonnoSegawa2005}.
The walk presented in the paper was operated by a Grover operator like Eq.~\eqref{eq:grover_coin}.
The choice of the operator enabled them to analyze the walk in a rigorous manner and two limit theorems for the position of the walker were proved.
One was a limit measure on the line and the other was a convergence in distribution in a rescaled space by time.
After a while, similar limit theorems for a three-state walk whose coin-flip operator contained the Grover operator were submitted~\cite{vStefavnakBezdvekovaJex2014,Machida2015}.
{\v{S}}tefa{\v{n}}{\'a}k et al.~\cite{vStefavnakBezdvekovaJex2014} examined a relation between limit distributions and particular initial conditions which came from the eigensystems of the coin-flip operator.
For a general initial condition at the origin, two limit theorems were computed by Machida~\cite{Machida2015}.
One of the features of the three-state quantum walk in the past studies was that the walker could localize.
Reading the proofs of the limit theorems, one can realize that a constant eigenvalue of an operator on a Fourier picture importantly works on localization.
The time-dependent three-state quantum walk defined in Sec.~\ref{sec:definition} can also localize due to a constant eigenvalue of an operator and that fact is expressed as the presence of a Dirac $\delta$-function in the following limit theorem.

\begin{thm}
For the time-dependent three-state quantum walk, there exists a non-negative constant $\Delta(\alpha,\beta,\gamma)$ whose value is determined by the values of the variables $\alpha,\beta$, and $\gamma\in\mathbb{C}$, and the cumulative distribution of the rescaled position $X_t/t$ converges in the form 
 \begin{align}
   &\lim_{t\to\infty}\mathbb{P}\left(\frac{X_t}{t}\leq x\right)\nonumber\\
  =&\int_{-\infty}^x \Bigl\{\Delta(\alpha,\beta,\gamma)\delta_0(y)+\nu(\alpha,\beta,\gamma; y)f(y)I_{\mathcal{D}_1}(y)+\nu(\gamma,\beta,\alpha; -y)f(-y)I_{\mathcal{D}_2}(y)\Bigr\}\,dy\quad (x\in\mathbb{R}),\label{eq:cumulative}
 \end{align}
 where $\delta_0(x)$ denotes a Dirac $\delta$-function at the origin and
 \begin{align}
  f(x)=&\frac{(1-c)\left\{5+4c-(4+5c)x^2+x\sqrt{D(x)}\right\}}{\pi(5+4c)^2(1-x^2)\sqrt{W_{+}(x)W_{-}(x)}\sqrt{D(x)}},\\[3mm]
  \nu(\alpha,\beta,\gamma; x)=&\xi_{-}(x)^2\left|\alpha\right|^2+\chi_1(x)^2\left|\beta\right|^2+\xi_{+}(x)^2\left|\gamma\right|^2\nonumber\\
  &-\xi_{-}(x)\chi_2(x)\Re(\alpha\overline{\beta})-\xi_{+}(x)\chi_2(x)\Re(\beta\overline{\gamma})+\left(\chi_1(x)^2-\frac{\chi_2(x)^2}{2}\right)\Re(\gamma\overline{\alpha}),\label{eq:function_nu}\\[3mm]
  D(x)=&(1-c)\left\{2(5+4c)-9(1+c)x^2\right\},\\
  W_{+}(x)=&1+2c-3cx^2+x\sqrt{D(x)},\\
  W_{-}(x)=&5+4c-3(2+c)x^2-x\sqrt{D(x)},\\[2mm]
  \xi_{+}(x)=&\frac{2(5+4c)+9(1+c)x+\sqrt{D(x)}}{2\sqrt{2}},\\
  \xi_{-}(x)=&\frac{-2(5+4c)+9(1+c)x+\sqrt{D(x)}}{2\sqrt{2}},\\
  \chi_1(x)=&\frac{3\left\{s^2x-(1+c)\sqrt{D(x)}\right\}}{2s},\\
  \chi_2(x)=&\frac{3(1-c)(4+3c)x-(2+c)\sqrt{D(x)}}{s},\\
  \mathcal{D}_1=&\left\{x\in\mathbb{R}\,\Big|\,-\frac{1+2c}{3} < x < \frac{\sqrt{5+4c}}{3}\right\},\\
  \mathcal{D}_2=&\left\{x\in\mathbb{R}\,\Big|\,-\frac{\sqrt{5+4c}}{3} < x < \frac{1+2c}{3}\right\},\\
  I_A(x)=&\left\{\begin{array}{cl}
	   1&(x\in A)\\
		  0&(x\notin A)
		 \end{array}\right..
 \end{align}
 Also, we have let $\Re(z)$ be the real part of a complex number $z$ in Eq.~\eqref{eq:function_nu}.
 \label{th:limit}
\end{thm}
\bigskip

Since the limit theorem states that the random variable $X_t$ rescaled by time $t$ (i.e. $X_t/t$) converges in distribution and the limit density function $(d/dx)\lim_{t\to\infty}\mathbb{P}(X_t/t\leq x)$, that is the integrand in Eq.~\eqref{eq:cumulative}, contains the continuous function $\nu(\alpha,\beta,\gamma; x)f(x)I_{\mathcal{D}_1}(x)+\nu(\gamma,\beta,\alpha; -x)f(-x)I_{\mathcal{D}_2}(x)$, we understand that the probability distribution spreads in proportion to time $t$.
We saw localization around the origin in the probability distribution $\mathbb{P}(X_t=x)$ in Figs.~\ref{fig:time-probability} and \ref{fig:theta-probability} and it is portrayed as the Dirac $\delta$-function at the origin on the rescaled line by time $t$. 
Also, the limit density function has a compact support as well as ones reported in the past studies and it is described by the sets $\mathcal{D}_1$ and $\mathcal{D}_2$.
Note that the inequality $|1+2\cos\theta|\leq \sqrt{5+4\cos\theta}$ holds for any $\theta\in [0,2\pi)$.
While the term $\Delta(\alpha,\beta,\gamma)\delta_0(x)$ in the limit density function catches localization of the probability distribution $\mathbb{P}(X_t=x)$ on the line $\mathbb{Z}$, the continuous part tells us an asymptotic behavior of ballistic spreading.
Consequently, the expression of the continuous part obtained in the theorem allows us to have an approximation of the probability $\mathbb{P}(X_t=x)$ at each position other than the origin in the form
\begin{align}
 &\mathbb{P}(X_t=x)\nonumber\\
 \sim &\frac{1}{t}\Bigl\{\nu(\alpha,\beta,\gamma; x/t)f(x/t)I_{\mathcal{D}_1}(x/t)+\nu(\gamma,\beta,\alpha; -x/t)f(-x/t)I_{\mathcal{D}_2}(x/t)\Bigr\} \quad (x=\pm 1, \pm 2,\ldots),\label{eq:approximation}
\end{align}
as time $t$ has gone enough up.
From the approximation, one can exactly estimate where the walker distributes except for around the origin, and it is answered as the region $(-\sqrt{5+4c}/3\,t, (1+2c)/3\,t) \cup (-(1+2c)/3\,t, \sqrt{5+4c}/3\,t)$.
Figure~\ref{fig:limit} illustrates the approximation in Eq.~\eqref{eq:approximation} and verifies the theorem when we set up $\alpha=\beta=\gamma=1/\sqrt{3}$.
As shown in Figs.~\ref{fig:limit}--(a) and (b), we see that the values given by the asymptotic function fit the ballistic spreading part of the probability distribution.
\bigskip

\begin{proof}{
The proof is demonstrated by using Fourier analysis which is a standard method to derive limit theorems in the field of quantum walks~\cite{GrimmettJansonScudo2004}.
Fourier analysis has played a role as a standard method to find limit theorems for quantum walks.
For the quantum walk defined in this paper, our analysis will be mainly concentrating on an eigenspace of a $3\times 3$ matrix.
Thanks to the existence of a constant eigenvalue, we can easily to understand that the quantum walk localizes in distribution.
The method, moreover, also has an advantage for the quantum walk.
We can reveal the possibility that it holds a gap in distribution, just by looking at a pair of functions.
With the method, we are approaching a convergence of the $r$-th moment $\mathbb{E}\left[(X_t/t)^r\right]\,(r\in\left\{0,1,2,\ldots\right\})$ as $t\to\infty$.
Let us define the Fourier transform of the quantum walk in the form
\begin{equation}
\ket{\hat\Psi_t(k)}=\sum_{x\in\mathbb{Z}}e^{-ikx}\ket{\psi_t(x)}\quad (k\in[-\pi,\pi)).
\end{equation}
Then, the superposition state all over the line is obtained from the inverse Fourier transform
\begin{equation}
 \ket{\Psi_t}=\sum_{x\in\mathbb{Z}}\ket{x}\otimes \int_{-\pi}^\pi e^{ikx}\ket{\hat\Psi_t(k)}\frac{dk}{2\pi},
\end{equation}
because one can extract the superposition state of the three coin states at position $x$ from an integral
\begin{equation}
 \ket{\psi_t(x)}=\int_{-\pi}^\pi e^{ikx}\ket{\hat\Psi_t(k)}\frac{dk}{2\pi}.
\end{equation}
The iteration defined in Eq.~(\ref{eq:time-evolution}) gives an update rule
\begin{equation}
 \ket{\hat\Psi_{t+1}(k)}=\left\{\begin{array}{ll}
			  \hat{S}(k)\hat{C}(k)\ket{\hat\Psi_t(k)}& (t=0,1 \mod 3)\\[1mm]
			   \hat{S}(k)\ket{\hat\Psi_t(k)}& (t=2 \mod 3)
				\end{array}\right.,
\end{equation}
from which
\begin{equation}
 \ket{\hat\Psi_{3t+\tau}(k)}=\hat C(k)^\tau\left(\hat S(k)\hat C(k)^2\right)^t\ket{\hat\Psi_0(k)} \quad (\tau=0,1,2),\label{eq:fourier_time}
\end{equation}
follows, where $\hat S(k)=e^{ik}\ket{-1}\bra{-1}+\ket{0}\bra{0}+e^{-ik}\ket{1}\bra{1}$ and $\hat C(k)=\hat S(k)C$.

From now on, we are describing the $r$-th moment $\mathbb{E}(X_t^r)$ in terms of the eigensystem of the operator $\hat S(k)\hat C(k)^2$.
Taking the vectors in the coin Hilbert space $\mathcal{H}_c$ to be
\begin{equation}
 \ket{-1}=\begin{bmatrix}
	   1\\0\\0 
	  \end{bmatrix},\quad	  
 \ket{0}=\begin{bmatrix}
	  0\\1\\0
	 \end{bmatrix},\quad
 \ket{1}=\begin{bmatrix}
	  0\\0\\1
	 \end{bmatrix},\label{eq:onb_H_c}
\end{equation}
we compute the eigenvalues and the eigenvectors of the operator.
The operator $\hat S(k)\hat C(k)^2$ is expressed as the matrix
\begin{equation}
 \begin{bmatrix}
  \frac{(1+c)^2e^{3ik}}{4}+\frac{s^2e^{2ik}}{2}+\frac{(1-c)^2e^{ik}}{4} & -\frac{(1+c)se^{3ik}}{2\sqrt{2}}+\frac{cse^{2ik}}{\sqrt{2}}+\frac{(1-c)se^{ik}}{2\sqrt{2}} & -\frac{s^2e^{3ik}}{4}+\frac{s^2e^{2ik}}{2}-\frac{s^2e^{ik}}{4}\\
  -\frac{(1+c)se^{ik}}{2\sqrt{2}}+\frac{cs}{\sqrt{2}}+\frac{(1-c)s}{2\sqrt{2}e^{ik}} & \frac{s^2e^{ik}}{2}+c^2+\frac{s^2}{2e^{ik}} & \frac{(1-c)se^{ik}}{2\sqrt{2}}+\frac{cs}{\sqrt{2}}-\frac{(1+c)s}{2\sqrt{2}e^{ik}}\\
  -\frac{s^2}{4e^{3ik}}+\frac{s^2}{2e^{2ik}}-\frac{s^2}{4e^{ik}} & -\frac{(1+c)s}{2\sqrt{2}e^{3ik}}+\frac{cs}{\sqrt{2}e^{2ik}}+\frac{(1-c)s}{2\sqrt{2}e^{ik}} & \frac{(1+c)^2}{4e^{3ik}}+\frac{s^2}{2e^{2ik}}+\frac{(1-c)^2}{4e^{ik}}
 \end{bmatrix},
\end{equation}
under the expression of Eq.~\eqref{eq:onb_H_c} and contains one constant eigenvalue which is independent from the variable $k$, and two eigenvalues which are functions of the variable $k$.
We let $\lambda_j(k)\,(j=1,2,3)$ be the eigenvalues and put
\begin{equation}
 \lambda_1(k)=1,\quad \lambda_j(k)=g(k)+(-1)^j i\sqrt{1-g(k)^2}\quad (j=2,3),\\
\end{equation}
where $i=\sqrt{-1}$ and
\begin{equation}
 g(k)=\frac{(1+c)^2}{4}\cos 3k+\frac{s^2}{2}\cos 2k+\frac{(1-c)(3+c)}{4}\cos k-\frac{s^2}{2}.
\end{equation}
Since the unitary matrix $\hat S(k)\hat C(k)^2$ contains a constant eigenvalue $\lambda_1(k)=1$, one can know that the walker holds the possibility of localization in distribution.
This fact will be shown in computation later.
The normalized eigenvector $\ket{v_j(k)}\,(j\in\left\{1,2,3\right\})$ associated to the eigenvalue $\lambda_j(k)$ has a form
\begin{equation}
 \ket{v_j(k)}=\frac{1}{\sqrt{N_j(k)}}\begin{bmatrix}
				      s(1-e^{ik})\left\{e^{ik}\eta(k)\lambda_j(k)-\eta(-k)\right\}\left\{\lambda_j(k)+1\right\}\\[2mm]
				      \sqrt{2}\left\{e^{ik}\eta(k)\lambda_j(k)-\eta(-k)\right\}\left\{e^{-ik}\eta(-k)\lambda_j(k)-\eta(k)\right\}\\[2mm]
					    s(1-e^{-ik})\left\{e^{-ik}\eta(-k)\lambda_j(k)-\eta(k)\right\}\left\{\lambda_j(k)+1\right\}				      
				     \end{bmatrix},
\end{equation}
where $\eta(k)=(1+c)e^{ik}+1-c$ and
\begin{align}
 N_{j}(k)=&\Bigl|s(1-e^{ik})\left\{e^{ik}\eta(k)\lambda_j(k)-\eta(-k)\right\}\left\{\lambda_j(k)+1\right\}\Bigr|^2\nonumber\\[2mm]
 &+\Bigl|\sqrt{2}\left\{e^{ik}\eta(k)\lambda_j(k)-\eta(-k)\right\}\left\{e^{-ik}\eta(-k)\lambda_j(k)-\eta(k)\right\}\Bigr|^2\nonumber\\[2mm]
 &+\Bigl|s(1-e^{-ik})\left\{e^{-ik}\eta(-k)\lambda_j(k)-\eta(k)\right\}\left\{\lambda_j(k)+1\right\}\Bigr|^2\quad (j\in\left\{1,2,3\right\}),
\end{align}
which denotes a normalization factor.
Since we have got the eigenvalues and the eigenvectors, we see the Fourier transform in the eigensystem of the operator $\hat S(k)\hat C(k)^2$,
\begin{equation}
 \ket{\hat\Psi_{3t+\tau}(k)}=\hat C(k)^\tau \sum_{j=1}^3\lambda_j(k)^t\braket{v_j(k)|\hat\Psi_0(k)}\ket{v_j(k)},
\end{equation}
using Eq.~\eqref{eq:fourier_time} and the decomposition $\ket{\hat\Psi_0(k)}=\sum_{j=1}^3\braket{v_j(k)|\hat\Psi_0(k)}\ket{v_j(k)}$.
Noting that the $r$-th derivative of the Fourier transform ($r\in\left\{0,1,2,\ldots\right\}$) has a representation
\begin{equation}
 (d/dk)^r\ket{\hat\Psi_{3t+\tau}(k)}=(t)_r\hat C(k)^\tau \sum_{j=1}^3\Bigl(\lambda'_j(k)^r\lambda_j(k)^{t-r}\braket{v_j(k)|\hat\Psi_0(k)}\ket{v_j(k)}\Bigr)+O(t^{r-1}),
\end{equation}
we get the $r$-th moment in an integral
\begin{align}
 \mathbb{E}(X_{3t+\tau}^r)=&\sum_{x\in\mathbb{Z}}x^r\mathbb{P}(X_{3t+\tau}=x)\nonumber\\
 =&\int_{-\pi}^\pi \bra{\hat\Psi_{3t+\tau}(k)}\left(D^r\ket{\hat\Psi_{3t+\tau}(k)}\right)\frac{dk}{2\pi}\nonumber\\
 =&(t)_r\int_{-\pi}^\pi \sum_{j=1}^3 \left(\frac{i\lambda'_j(k)}{\lambda_j(k)}\right)^r\left|\braket{v_j(k)|\hat\Psi_0(k)}\right|^2\frac{dk}{2\pi}+O(t^{r-1}),
 \label{eq:r-th_moment}
\end{align}
where $D=i(d/dk)$ and $(t)_r=\Pi_{j=1}^{r}\, (t-r+j)=t(t-1)\times\cdots\times(t-r+1)$.
Moving our focuses to the $r$-th moment in a rescaled space by time,
\begin{align}
 &\mathbb{E}\left[\left(\frac{X_{3t+\tau}}{3t+\tau}\right)^r\right]=\frac{\mathbb{E}(X_{3t+\tau})^r}{(3t)^r}\cdot \left(\frac{3t}{3t+\tau}\right)^r\nonumber\\
 =&\left\{\frac{(t)_r}{t^r}\int_{-\pi}^\pi \sum_{j=1}^3 \left(\frac{i\lambda'_j(k)}{3\lambda_j(k)}\right)^r\left|\braket{v_j(k)|\hat\Psi_0(k)}\right|^2\frac{dk}{2\pi}+\frac{O(t^{r-1})}{(3t)^r}\right\}\cdot \left(\frac{3t}{3t+\tau}\right)^r,
\end{align}
we realize that it is approaching a value
\begin{equation}
 \sum_{j=1}^3 \int_{-\pi}^\pi \left(\frac{i\lambda'_j(k)}{3\lambda_j(k)}\right)^r\left|\braket{v_j(k)|\hat\Psi_0(k)}\right|^2\frac{dk}{2\pi},
\end{equation}
as $t\to\infty$.
We should note that the limit has nothing to do with the value of the variable $\tau\in\left\{0,1,2\right\}$.
It turns out that the $r$-th moment $\mathbb{E}[(X_t/t)^r]$ has the limit as $t\to\infty$, and its value can be expressed as an integral,
\begin{equation}
  \lim_{t\to\infty}\mathbb{E}\left[\left(\frac{X_t}{t}\right)^r\right]=\sum_{j=1}^3 \int_{-\pi}^\pi \left(\frac{i\lambda'_j(k)}{3\lambda_j(k)}\right)^r\left|\braket{v_j(k)|\hat\Psi_0(k)}\right|^2\frac{dk}{2\pi}.\label{eq:r-th_moment_integral}
\end{equation}
The constant eigenvalue $\lambda_1(k)=1$ slightly changes Eq.~\eqref{eq:r-th_moment_integral} into
\begin{equation}
 \lim_{t\to\infty}\mathbb{E}\left[\left(\frac{X_t}{t}\right)^r\right]=0^r\cdot\Delta(\alpha,\beta,\gamma)+\sum_{j=2}^3 \int_{-\pi}^\pi\left(\frac{i\lambda'_j(k)}{3\lambda_j(k)}\right)^r\left|\braket{v_j(k)|\hat\Psi_0(k)}\right|^2\frac{dk}{2\pi},\label{eq:150822_4}
\end{equation}
where
\begin{equation}
 \Delta(\alpha,\beta,\gamma)=\int_{-\pi}^\pi\left|\braket{v_1(k)|\hat\Psi_0(k)}\right|^2\frac{dk}{2\pi}.\label{eq:Delta}
\end{equation}
Although Theorem~\ref{th:limit} has asserted that the constant $\Delta(\alpha,\beta,\gamma)$ takes a non-negative value, it is actually greater than 0.
That fact will be shown after this proof.
On the other hand, the function $i\lambda'_j(k)/3\lambda_j(k)\,(j\in\left\{2,3\right\})$ is of the form
\begin{equation}
 \frac{i\lambda'_j(k)}{3\lambda_j(k)}=(-1)^j\frac{\left\{c-(1+c)\cos k\right\}\sin k}{\bigl|\left\{c-(1+c)\cos k\right\}\sin k\bigr|}\cdot \frac{2+c+3(1+c)\cos k}{3\sqrt{2-c^2+2(1+c)\cos k+(1+c)^2\cos^2 k}}.
\end{equation}
Seeing two examples of the functions $i\lambda'_j(k)/3\lambda_j(k)\,(j=2,3)$ in Figure~\ref{fig:h(k)}, one can be easily aware of a possible gap in distribution.
Indeed, when $\theta=5\pi/6$, the range of the functions out of  the interval $((1+2c)/3,\,-(1+2c)/3)$, as shown in Fig.~\ref{fig:h(k)}-(b).
We should note that $1+2c=1+2\cos(5\pi/6)=-0.732\cdots<0$ in Fig.~\ref{fig:h(k)}-(b).
Just examining the functions $i\lambda'_j(k)/3\lambda_j(k)\,(j=2,3)$, we have realized the possibility of a gap for the 3-period time-dependent quantum walk defined in this paper.
\begin{figure}[h]
\begin{center}
 \begin{minipage}{70mm}
  \begin{center}
   \includegraphics[scale=0.5]{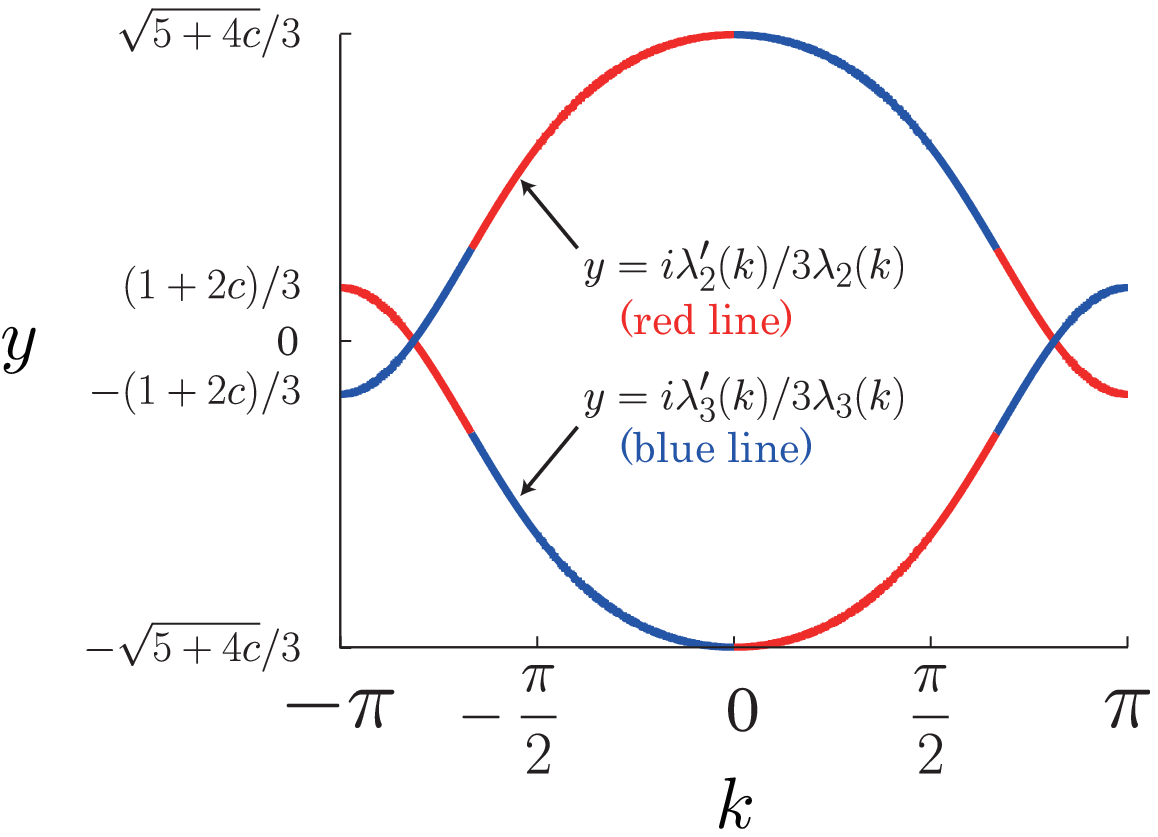}\\[2mm]
  (a) $\theta=\arccos(-1/3)$
  \end{center}
 \end{minipage}
 \begin{minipage}{70mm}
  \begin{center}
   \includegraphics[scale=0.5]{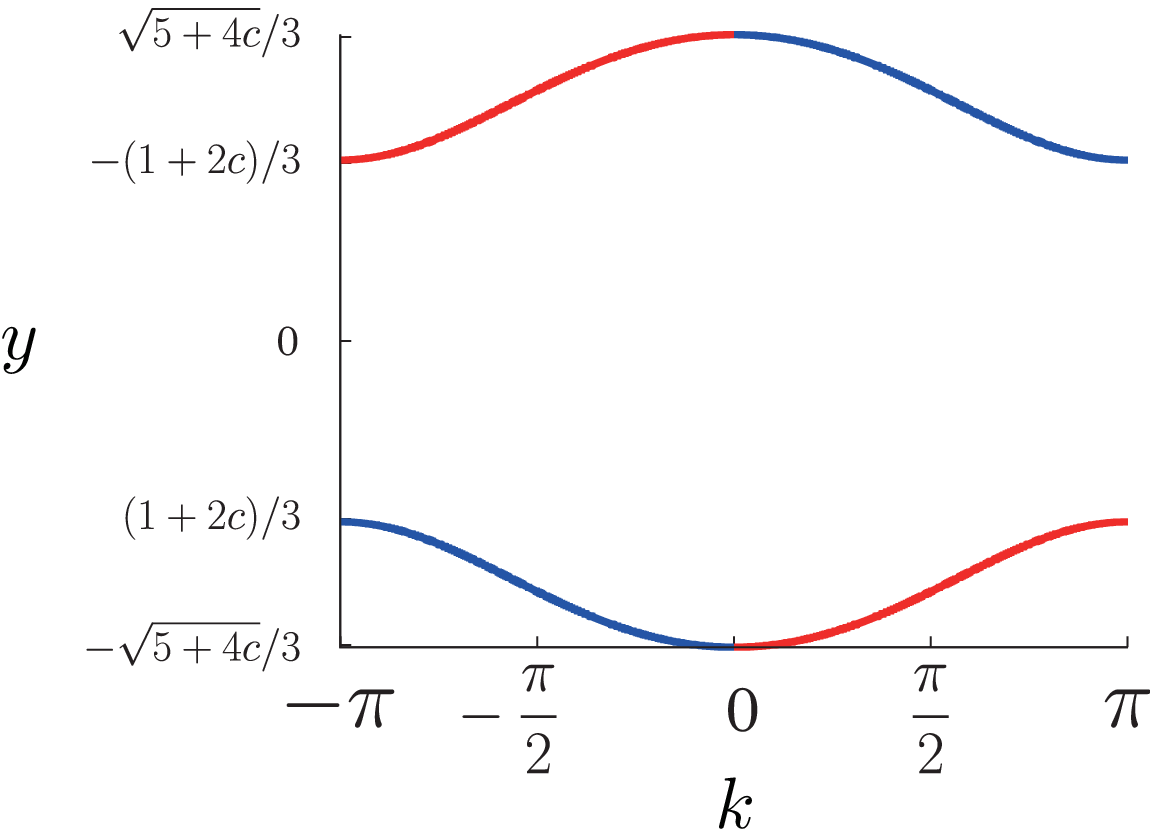}\\[2mm]
  (b) $\theta=5\pi/6$
  \end{center}
 \end{minipage}
\vspace{5mm}
\fcaption{The graphs of $y=i\lambda'_2(k)/\lambda_2(k)$ (red line) and $y=i\lambda'_3(k)/\lambda_3(k)$ (blue line) : Figure-(b) shows that the functions do not take any value in the range $((1+2c)/3,\,-(1+2c)/3)$, which means that the quantum walker gives rise to a gap around the origin in distribution when $\theta=5\pi/6$.}
\label{fig:h(k)}
\end{center}
\end{figure}

Substitutions $i\lambda'_j(k)/(3\lambda_j(k))=x\,(j=2,3)$ in Eq.~(\ref{eq:150822_4}) produce
\begin{align}
 &\lim_{t\to\infty}\mathbb{E}\left[\left(\frac{X_t}{t}\right)^r\right]\nonumber\\
 =&\int_{-\infty}^\infty x^r \Bigl\{\Delta(\alpha,\beta,\gamma)\delta_0(x)+\nu(\alpha,\beta,\gamma; x)f(x)I_{\mathcal{D}_1}(x)+\nu(\gamma,\beta,\alpha; -x)f(-x)I_{\mathcal{D}_2}(x)\Bigr\}\,dx.\label{eq:limit_moment}
\end{align}
The right hand side of Eq.~\eqref{eq:limit_moment} shows the $r$-th moment of a random variable whose density function is the integrand in Eq.~\eqref{eq:cumulative}.
At the end, a family of the limits $\lim_{t\to\infty}\mathbb{E}[(X_t/t)^r]\,(r=0,1,2,\ldots)$ means that the rescaled position $X_t/t$ converges to the random variable in distribution, and one can conclude Eq.~\eqref{eq:cumulative}. 
}
\end{proof}

\bigskip

Now, we discuss localization of the quantum walk presented in this paper.
A chance of localization is given by both the time evolution operators and the initial condition.
Since we already saw that the operator $\hat S(k)\hat C(k)^2$ had a constant eigenvalue $\lambda_1(k)=1$ which made the Dirac $\delta$-function $\delta_0(x)$ in the limit distribution, there is a possibility that the walker localizes, as described by the term $\Delta(\alpha,\beta,\gamma)\delta_0(x)$.
Again, we should note that localization also depends on the initial condition.
Instead of localization, one may focus on delocalization.
To discuss what initial condition causes delocalization, let us try to find the values of the variables $\alpha,\beta$, and $\gamma\in\mathbb{C}$ such that
\begin{equation}
 \Delta(\alpha,\beta,\gamma)=\int_{-\pi}^{\pi}\Bigl|\braket{v_1(k)|\hat\Psi_0(k)}\Bigr|^2\,\frac{dk}{2\pi}=0.\label{eq:assumption}
\end{equation}
If we find such values, then the walker delocalizes in distribution for them due to the lack of the Dirac $\delta$-function.
But, there do not exist such values actually, as shown below.
Assuming Eq.~\eqref{eq:assumption}, we begin to prove that fact.
Since we estimate
\begin{equation}
 \int_{-\pi}^\pi\Bigl|\braket{v_1(k)|\hat\Psi_0(k)}\Bigr|\,\frac{dk}{2\pi} \leq \left(\int_{-\pi}^\pi\Bigl|\braket{v_1(k)|\hat\Psi_0(k)}\Bigr|^2\,\frac{dk}{2\pi}\right)^{1/2},
\end{equation}
from H\"{o}lder's inequality, a condition
\begin{equation}
 \int_{-\pi}^\pi\Bigl|\braket{v_1(k)|\hat\Psi_0(k)}\Bigr|\,\frac{dk}{2\pi}=0,\label{eq:150822_3}
\end{equation}
is required to hold.
Multiplying both sides of Eq.~\eqref{eq:150822_3} by $\sup_{k\in[-\pi,\pi)}\sqrt{N_1(k)}\cdot 2\pi$ guarantees
\begin{equation}
 \int_{-\pi}^\pi\Bigl|\braket{v_1(k)|\hat\Psi_0(k)}\Bigr|\cdot\sup_{k\in[-\pi,\pi)}\sqrt{N_1(k)}\,dk=0,
\end{equation}
from which we get
\begin{equation}
 \int_{-\pi}^\pi\Bigl|\sqrt{N_1(k)}\braket{v_1(k)|\hat\Psi_0(k)}\Bigr|\,dk =0,\label{eq:150822_1}
\end{equation}
because of an inequality
\begin{equation}
 \int_{-\pi}^\pi\Bigl|\sqrt{N_1(k)}\braket{v_1(k)|\hat\Psi_0(k)}\Bigr|\,dk \leq \int_{-\pi}^\pi\Bigl|\braket{v_1(k)|\hat\Psi_0(k)}\Bigr|\cdot\sup_{k\in[-\pi,\pi)}\sqrt{N_1(k)}\,dk.
\end{equation}
On the other hand, we also have an inequality  
\begin{align}
 \forall y\in\mathbb{Z},\quad \left|\int_{-\pi}^\pi e^{iky}\sqrt{N_1(k)}\braket{v_1(k)|\hat\Psi_0(k)}\,dk\right| &\leq \int_{-\pi}^\pi \Bigl|e^{iky}\sqrt{N_1(k)}\braket{v_1(k)|\hat\Psi_0(k)}\Bigr|\,dk\nonumber\\
 &=\int_{-\pi}^\pi \Bigl|\sqrt{N_1(k)}\braket{v_1(k)|\hat\Psi_0(k)}\Bigr|\,dk.\label{eq:150822_2}
\end{align}
Equations~\eqref{eq:150822_1} and \eqref{eq:150822_2} give a condition for us to judge delocalization,
\begin{equation}
 \forall y\in\mathbb{Z},\quad \int_{-\pi}^\pi e^{iky}\sqrt{N_1(k)}\braket{v_1(k)|\hat\Psi_0(k)}\,dk=0,
\end{equation}
and the representation
\begin{align}
 \sqrt{N_1(k)}\braket{v_1(k)|\hat\Psi_0(k)}=&-e^{-3ik}\cdot (1+c)\left\{\sqrt{2}s\alpha+(1+c)\beta\right\}+e^{-2ik}\cdot 2s\left(\sqrt{2}c\alpha-s\beta\right)\nonumber\\
 &+e^{-ik}\cdot\left\{2\sqrt{2}(1-c)s\alpha+(1-c)(1+3c)\beta-\sqrt{2}(1+c)s\gamma\right\}\nonumber\\
 &+2\sqrt{2}\left\{cs\alpha+\sqrt{2}(1+c^2)\beta+cs\gamma\right\}\nonumber\\
 &+e^{ik}\cdot\left\{-\sqrt{2}(1+c)s\alpha+(1-c)(1+3c)\beta+2\sqrt{2}(1-c)s\gamma\right\}\nonumber\\
 &+e^{2ik}\cdot2s\left(-s\beta+\sqrt{2}c\gamma\right)-e^{3ik}\cdot(1+c)\left\{(1+c)\beta+\sqrt{2}s\gamma\right\},
\end{align}
allows us to hold the conditions which the variables $\alpha,\beta$, and $\gamma\in\mathbb{C}$ should satisfy,
\begin{align}
 \sqrt{2}s\alpha+(1+c)\beta&=0,\\
 \sqrt{2}c\alpha-s\beta&=0,\\
 2\sqrt{2}(1-c)s\alpha+(1-c)(1+3c)\beta-\sqrt{2}(1+c)s\gamma&=0,\\
 cs\alpha+\sqrt{2}(1+c^2)\beta+cs\gamma&=0,\\
 -\sqrt{2}(1+c)s\alpha+(1-c)(1+3c)\beta+2\sqrt{2}(1-c)s\gamma&=0,\\
 -s\beta+\sqrt{2}c\gamma&=0,\\
 (1+c)\beta+\sqrt{2}s\gamma&=0.
\end{align}
However, there do not exist complex numbers $\alpha,\beta$, and $\gamma$ which satisfy both these conditions and $|\alpha|^2+|\beta|^2+|\gamma|^2=1$, and then one can understand that the coefficient $\Delta(\alpha,\beta,\gamma)$ of the Dirac $\delta$-function is not equal to but greater than 0 for any values of the variables $\alpha, \beta$, and $\gamma$ such that $|\alpha|^2+|\beta|^2+|\gamma|^2=1$.
Consequently, the walker certainly localizes at the origin in the rescaled space by time $t$ due to the presence of the Dirac $\delta$-function $\delta_0(x)$ in the limit density function.

\section{Summary and Discussion}
\label{sec:summary}
The time-dependent three-state quantum walk presented in this paper showed both localization and ballistic spreading in distribution.
We viewed some configurations of the walker in Figs.~\ref{fig:time-probability} and \ref{fig:theta-probability}, and confirmed them in a long-time limit theorem as $t\to\infty$.
The limit theorem told us that the walker was spreading in proportion to time $t$ and localized in distribution.
For each time $t$, the ballistic spreading part was given on a region which was described as a compact support in a rescaled space by time $t$, and we observed a gap in it for suitable values of the parameter $\theta$ which we used to build the coin-flip operator $C$ in Eq.~\eqref{eq:coin-flip_operator}.
If we picked a value out of the interval $(2\pi/3,4\pi/3)$ and substituted it to the parameter $\theta$, then the continuous part of the limit density function returned the value 0 on the interval $((1+2\cos\theta)/3,-(1+2\cos\theta)/3)$ because of $1+2\cos\theta<0$ for $\theta\in(2\pi/3,4\pi/3)$, which meant the birth of a gap in distribution.
That fact was completely reproduced in Fig.~\ref{fig:limit} by using an approximate function in Eq.~\eqref{eq:approximation} obtained from the limit theorem.
Although a distribution with a gap was already reported for a 3-period time-dependent two-state quantum walk in 2015, the two-state walk did not localize at all due to the lack of a constant eigenvalue of its time evolution operator on a Fourier picture~\cite{GrunbaumMachida2015}.
One can say that an existence of a probability distribution which holds both localization and a gap has been discovered, and we have never seen such an interesting distribution in the field of quantum walks before.
  
We also proved that the walker certainly localized for any values of the variables $\alpha, \beta$, and $\gamma\in\mathbb{C}$ which determined the initial condition of the walker.
Whereas, for the standard (time-independent) three-state quantum walks defined in the past studies~\cite{InuiKonnoSegawa2005,Machida2015}, one can give delocalization to the quantum walks using a suitable initial condition.
That fact should be a highlighted difference between the standard walks and the time-dependent walk examined in this paper.
The strength of localization was measured in the coefficient of the Dirac $\delta$-function in the limit density function.
The coefficient, however, still remains to be given in the form of a definite integral, as written in Eq.~\eqref{eq:Delta}, due to a difficulty of evaluating it.
It should be a future problem to make the behavior of localization clearer by finding the value of the integral.

\nonumsection{Acknowledgements}
\noindent The author is grateful to the Japan Society for the Promotion of Science for the support.

\appendix

We analyzed a quantum walk on which coin operator $C$ did not work at time $t=2 \mod 3$.
Is this appendix we briefly observe the quantum walker lacking the coin operator at time $t=0 \mod 3$ or $t=1 \mod 3$.
Each quantum walk can be treated the same as a quantum walk whose dynamics is given by Eq.~\eqref{eq:time-evolution} with a delocalized initial state.
We will, moreover, see their probability distributions in numerics.

\begin{figure}[h]
\begin{center}
 \begin{minipage}{70mm}
  \begin{center}
   \includegraphics[scale=0.5]{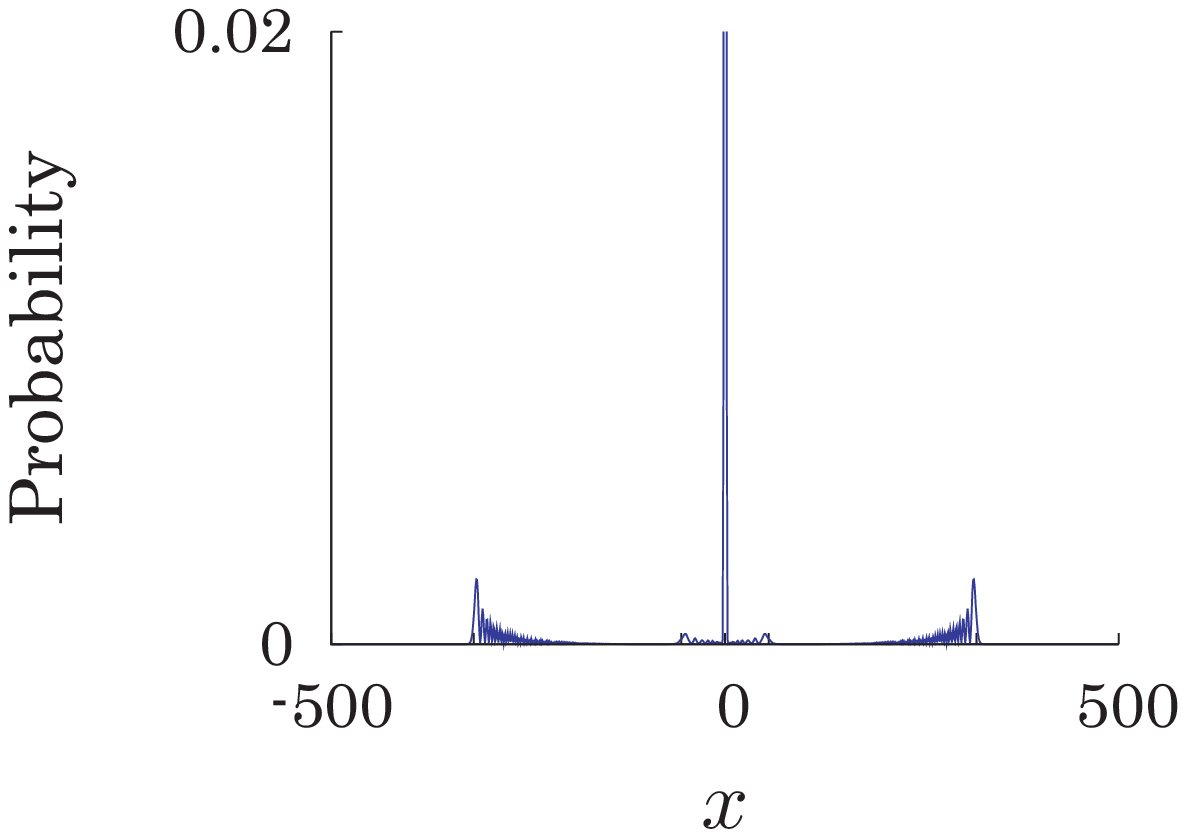}\\[2mm]
  (a) $\theta=\arccos(-1/3)$
  \end{center}
 \end{minipage}
 \begin{minipage}{70mm}
  \begin{center}
   \includegraphics[scale=0.5]{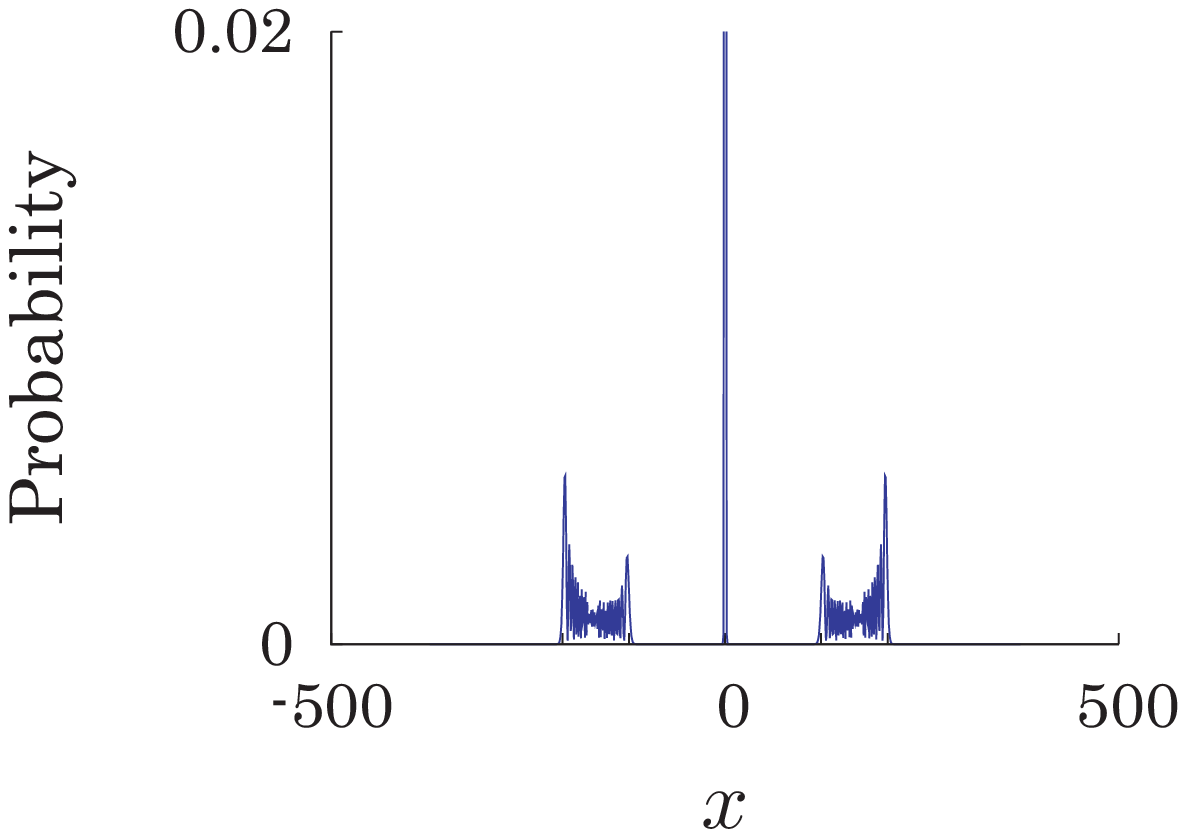}\\[2mm]
  (b) $\theta=5\pi/6$
  \end{center}
 \end{minipage}
\vspace{5mm}
\fcaption{Probability distribution at time $500$ for the quantum walk lacking coin operator $C$ at time $t=0 \mod 3$}
\label{fig:case_1}
\end{center}
\end{figure}

First, given a localized initial state by Eq.~\eqref{eq:initial_state}, the quantum walk is updating without coin operator $C$ at time $t=0 \mod 3$,
\begin{equation}
 \ket{\Psi_{t+1}}=\left\{\begin{array}{ll}
		   \tilde{S}\tilde{C}\ket{\Psi_t}& (t=1,2 \mod 3)\\[1mm]
			  \tilde{S}\ket{\Psi_t}& (t=0 \mod 3)
			 \end{array}\right..
\end{equation}
Since $\ket{\Psi_1}=\tilde{S}\ket{\Psi_0}$, this quantum walk at time $t\,(t=1,2,\ldots)$ is equivalent in distribution to the quantum walk at time $t-1$ whose evolution is given by Eq.~\eqref{eq:time-evolution} with a delocalized initial state
\begin{equation}
 \ket{\Psi_0}=\tilde{S}\left\{\ket{0}\otimes\left(\alpha\ket{-1}+\beta\ket{0}+\gamma\ket{1}\right)\right\}=\alpha\ket{-1}\otimes\ket{-1}+\beta\ket{-1}\otimes\ket{0}+\gamma\ket{1}\otimes\ket{1}.
\end{equation}
As shown in Fig.~\ref{fig:case_1}, localization and a gap can appear on the quantum walk in distribution.

\begin{figure}[h]
\begin{center}
 \begin{minipage}{70mm}
  \begin{center}
   \includegraphics[scale=0.5]{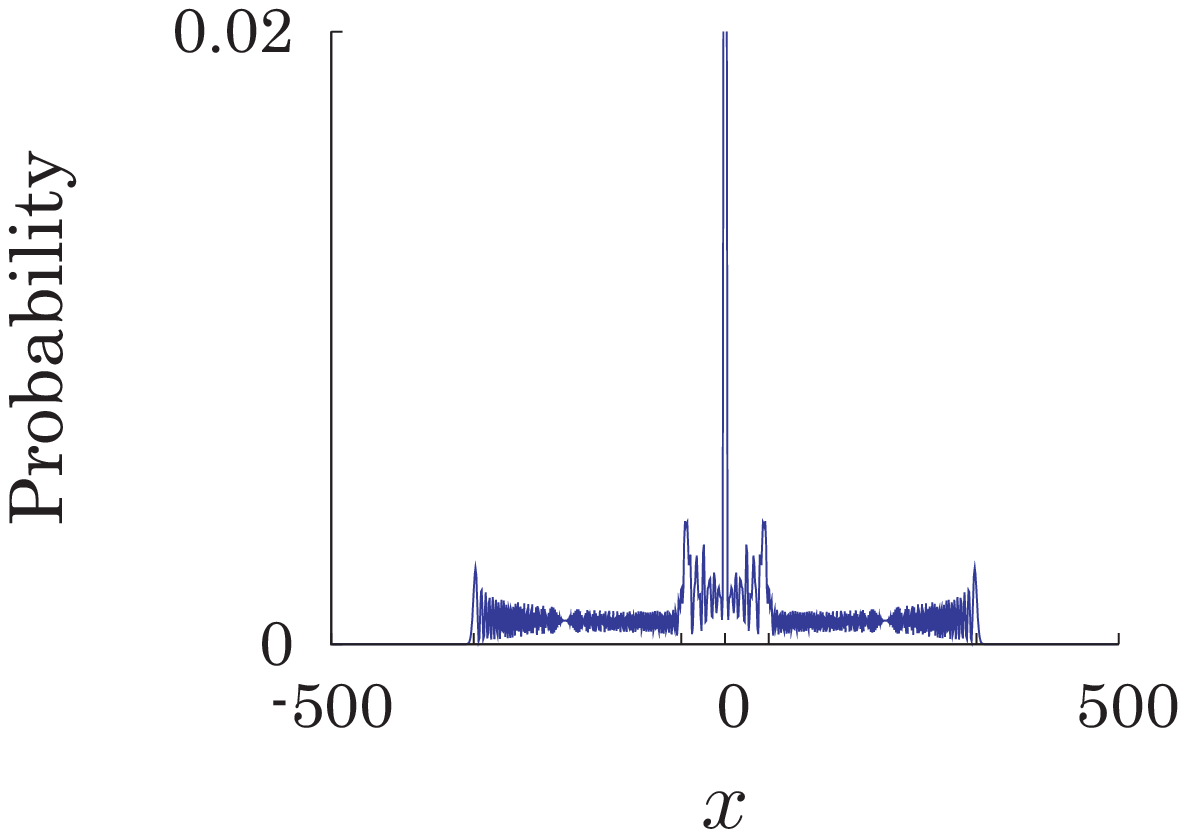}\\[2mm]
  (a) $\theta=\arccos(-1/3)$
  \end{center}
 \end{minipage}
 \begin{minipage}{70mm}
  \begin{center}
   \includegraphics[scale=0.5]{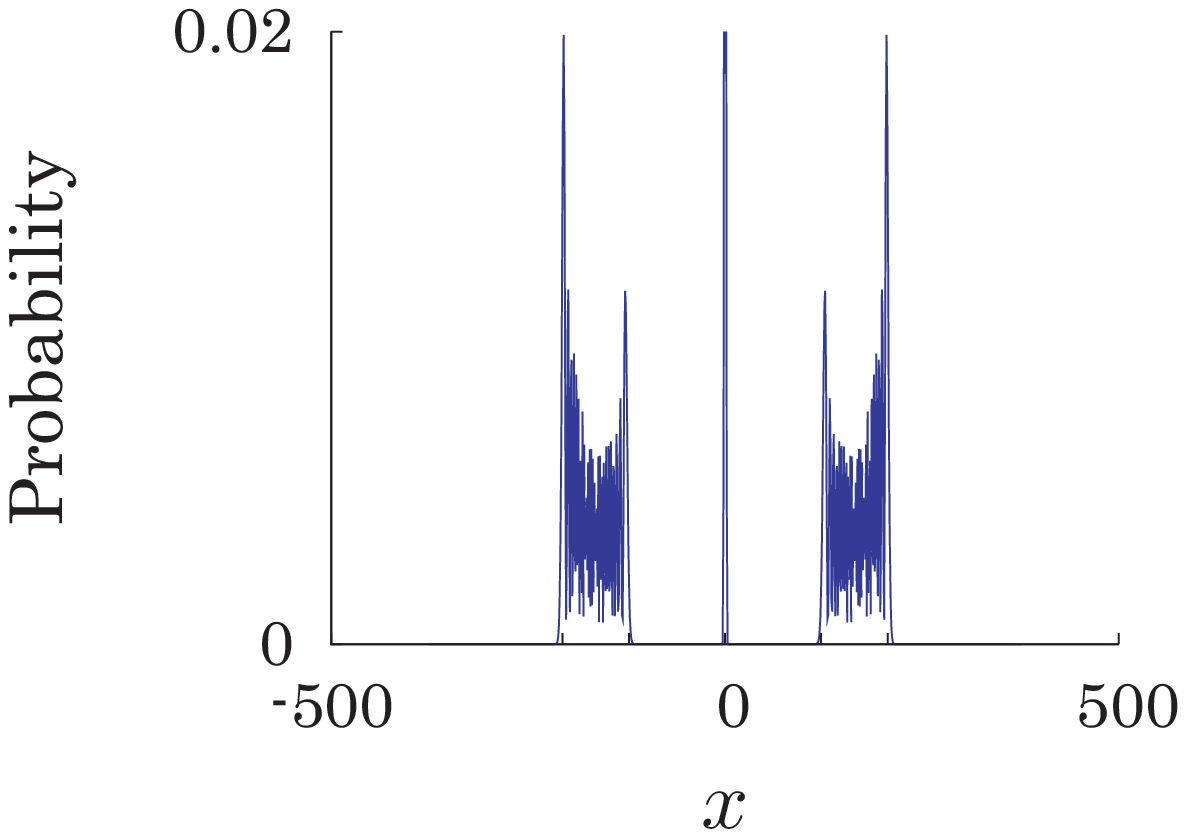}\\[2mm]
  (b) $\theta=5\pi/6$
  \end{center}
 \end{minipage}
\vspace{5mm}
\fcaption{Probability distribution at time $500$ for the quantum walk lacking coin operator $C$ at time $t=1 \mod 3$}
\label{fig:case_2}
\end{center}
\end{figure}

On the other hand, if a localized initial state Eq.~\eqref{eq:initial_state} is allocated to the quantum walker and it repeats the iteration
\begin{equation}
 \ket{\Psi_{t+1}}=\left\{\begin{array}{ll}
		   \tilde{S}\tilde{C}\ket{\Psi_t}& (t=0,2 \mod 3)\\[1mm]
			  \tilde{S}\ket{\Psi_t}& (t=1 \mod 3)
			 \end{array}\right.,
\end{equation}
then probability distribution $\mathbb{P}(X_t=x)$ can show localization and a gap again, as viewed in Fig.~\ref{fig:case_2}.
From the fact $\ket{\Psi_2}=\tilde{S}(\tilde{S}\tilde{C})\ket{\Psi_0}$, this quantum walk at time $t\,(t=2,3,\ldots)$ is same as the walk at time $t-2$ whose evolution is defined by Eq.~\eqref{eq:time-evolution} starting off with a delocalized initial state
\begin{align}
 \ket{\Psi_0}=&\tilde{S}(\tilde{S}\tilde{C})\left\{\ket{0}\otimes\left(\alpha\ket{-1}+\beta\ket{0}+\gamma\ket{1}\right)\right\}\nonumber\\
 =&\left(-\frac{1+c}{2}\alpha+\frac{s}{\sqrt{2}}\beta+\frac{1-c}{2}\gamma\right)\ket{-2}\otimes\ket{-1}
 +\left(\frac{s}{\sqrt{2}}\alpha+c\beta+\frac{s}{\sqrt{2}}\gamma\right)\ket{0}\otimes\ket{0}\nonumber\\
 &+\left(\frac{1-c}{2}\alpha+\frac{s}{\sqrt{2}}\beta-\frac{1+c}{2}\gamma\right)\ket{2}\otimes\ket{1}.
\end{align}

\bigskip


\end{document}